\documentclass[%
 reprint,
superscriptaddress,
 amsmath,amssymb,
 aps,
 pra, longbibliography
]{revtex4-1}

\usepackage{graphicx}% Include figure files
\usepackage{dcolumn}% Align table columns on decimal point
\usepackage{bm}% bold math
\usepackage{epstopdf}
\usepackage[usenames,dvipsnames]{xcolor}
\usepackage{gensymb}
\begin{document}

%\preprint{APS/123-QED}

\title{Hybridization and spin-orbit coupling effects\\ in quasi-one-dimensional spin-$\frac12$ magnet Ba$_{3}$Cu$_{3}$Sc$_{4}$O$_{12}$}% Force line breaks with \\
%\thanks{A footnote to the article title}%
\author{D.I. Badrtdinov}
\affiliation{Theoretical Physics and Applied Mathematics Department, Ural Federal University, 620002 Ekaterinburg, Russia}

\author{O.S. Volkova}
\affiliation{Theoretical Physics and Applied Mathematics Department, Ural Federal University, 620002 Ekaterinburg, Russia}
\affiliation{Low Temperature Physics and Superconductivity Department, Moscow State University, Moscow 119991, Russia}

\author{A.A. Tsirlin}
\affiliation{Experimental Physics VI, Center for Electronic Correlations and Magnetism, Institute of Physics, University of Augsburg, 86135 Augsburg, Germany}

\author{I.V. Solovyev}
\affiliation{Theoretical Physics and Applied Mathematics Department, Ural Federal University, 620002 Ekaterinburg, Russia}
\affiliation{Computational Materials Science Unit, National Institute for Materials Science, Tsukuba 305-0044, Japan}

\author{A.N. Vasiliev}
\affiliation{Theoretical Physics and Applied Mathematics Department, Ural Federal University, 620002 Ekaterinburg, Russia}
\affiliation{Low Temperature Physics and Superconductivity Department, Moscow State University, Moscow 119991, Russia}

\author{V.V. Mazurenko}
\affiliation{Theoretical Physics and Applied Mathematics Department, Ural Federal University, 620002 Ekaterinburg, Russia}

\date{\today}% It is always \today, today,
             %  but any date may be explicitly specified

\begin{abstract}
We study electronic and magnetic properties of the quasi-one-dimensional spin-$\frac12$ magnet Ba$_{3}$Cu$_{3}$Sc$_{4}$O$_{12}$ with a distinct orthogonal connectivity of CuO$_4$ plaquettes. An effective low-energy model taking into account spin-orbit coupling was constructed by means of first-principles calculations. On this basis a complete microscopic magnetic model of Ba$_{3}$Cu$_{3}$Sc$_{4}$O$_{12}$, including symmetric and antisymmetric anisotropic exchange interactions, is derived. The anisotropic exchanges are obtained from a distinct first-principles numerical scheme combining, on one hand, the local density approximation taking into account spin-orbit coupling, and, on the other hand, projection procedure along with the microscopic theory by Toru Moriya. The resulting tensors of the symmetric anisotropy favor collinear magnetic order along the structural chains with the leading ferromagnetic coupling $J_1\simeq -9.88$\,meV. The interchain interactions $J_8\simeq 0.21$\,meV and $J_5\simeq 0.093$\,meV are antiferromagnetic.  Quantum Monte Carlo simulations demonstrated that the proposed model reproduces the experimental Neel temperature, magnetization and magnetic susceptibility data. The modeling of neutron diffraction data reveals an important role of the covalent Cu--O bonding in Ba$_{3}$Cu$_{3}$Sc$_{4}$O$_{12}$. 
\end{abstract}

\maketitle

%\tableofcontents

\section{\label{sec:level1}Introduction}
The CuO$_4$ plaquette is the key structural element in the majority of low-dimensional copper oxides~\cite{Guerrero,Tsirlin,Janson,Vasiliev}. Ideally, it consists of the Cu atom at the center of the square formed by four oxygen atoms. The valence 2+ of the Cu ion corresponds to the atomic configuration in which all $3d$ orbitals are fully occupied except that of the $x^2-y^2$ symmetry containing one unpaired electron and having the highest energy. Such a geometry leads to a strong hybridization between the $x^2-y^2$ states of Cu and $2p$ states of oxygen, which affects magnetic properties of low-dimensional cuprates. For instance, spin density is largely delocalized and spread over the plaquette. This effect should be taken into account when analyzing the data from neutron scattering~\cite{FormFactor}.

The mutual orientation of the CuO$_4$ plaquettes in a particular system is very important and responsible for nontrivial magnetic and electronic properties observed experimentally. Depending on the connectivity of the plaquettes, the interaction between magnetic moments can be ferromagnetic or antiferromagnetic. Here, the copper-oxygen-copper bond angle plays crucial role.

\begin{figure}[!h]
\includegraphics[width=0.23\textwidth]{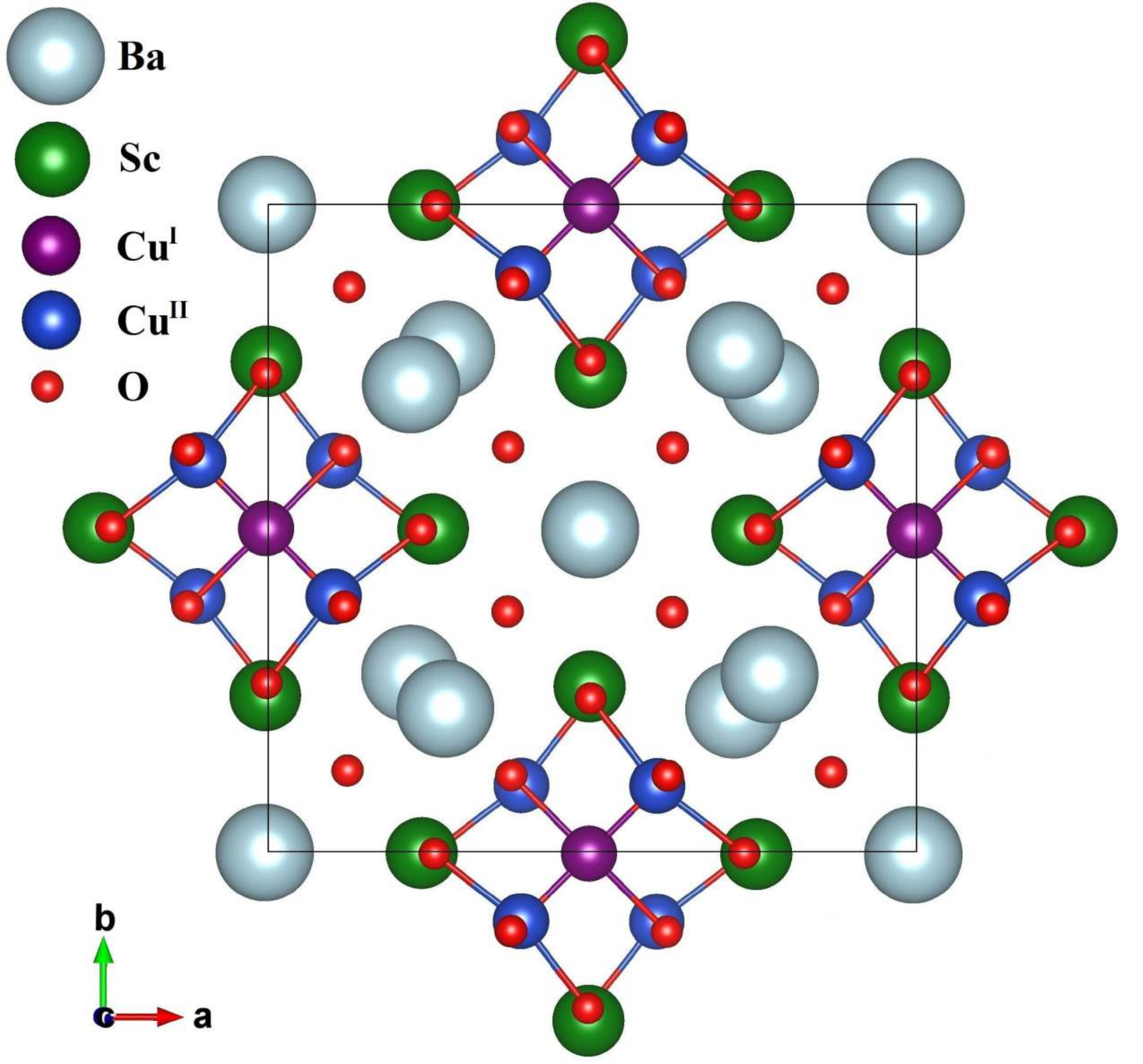}
\includegraphics[width=0.23\textwidth]{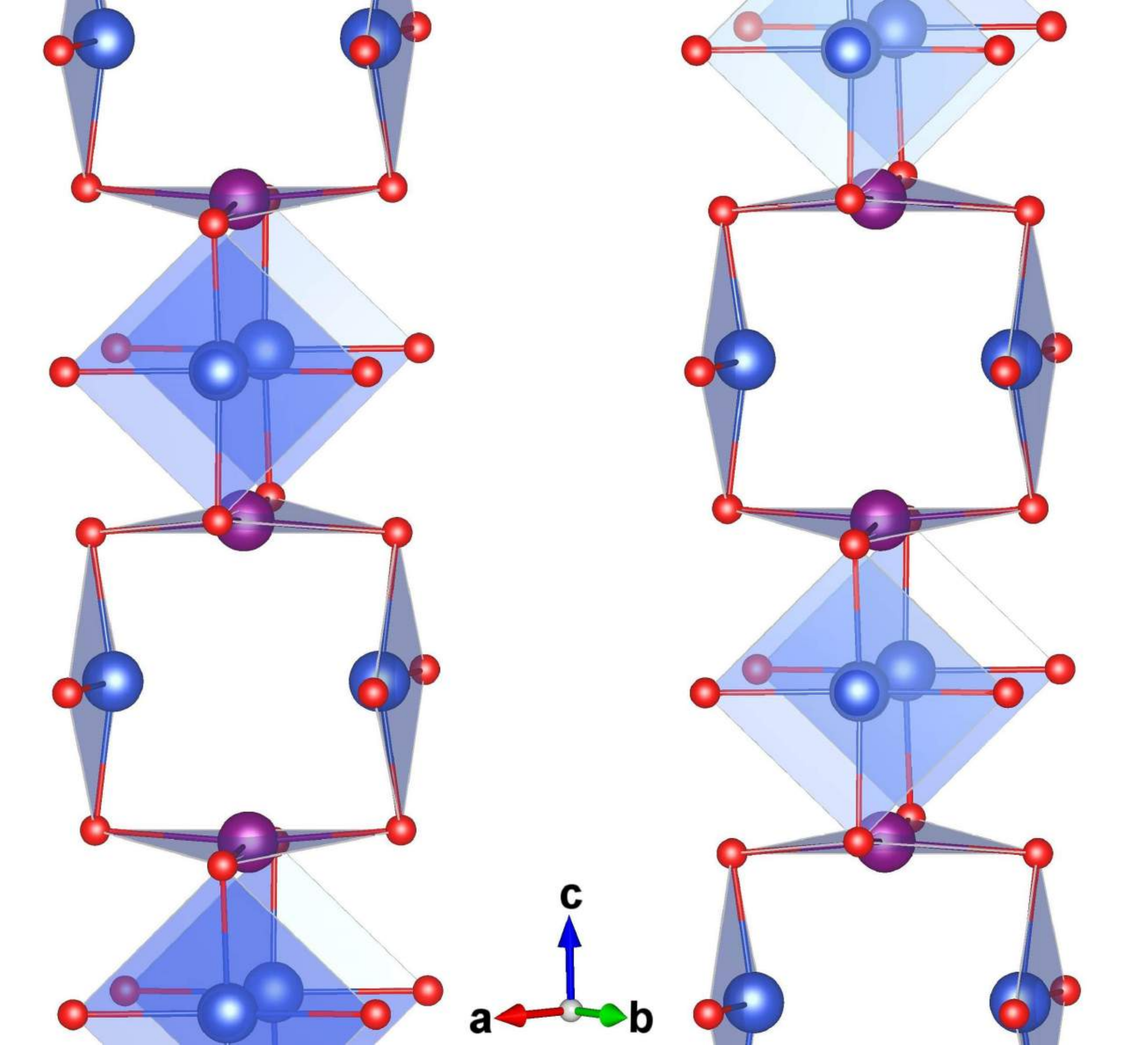}
\caption{(Left panel) Crystal structure of Ba$_{3}$Cu$_{3}$Sc$_{4}$O$_{12}$. (Right panel) Paper chains formed by CuO$_4$ plaquettes. Barium, scandium and interchain oxygen atoms are excluded for clarity. Crystal structures are visualized by using the VESTA software~\cite{VESTA}.}
\label{im:Crystal}
\end{figure}

The topology of Ba$_{3}$Cu$_{3}$Sc$_{4}$O$_{12}$ and isostructural Ba$_3$Cu$_3$In$_4$O$_{12}$ is a distinct one, since it contains three mutually orthogonal plaquette sublattices presented in Fig.~\ref{im:Crystal}. This type of geometry was coined a paper-chain model~\cite{Volkova}. Within each paper chain, the neighboring CuO$_4$ plaquettes have one common oxygen site. The corresponding angle of the Cu--O--Cu bond is close to 90$^{\circ}$, which means that kinetic superexchange processes between the neighboring Cu sites in Ba$_{3}$Cu$_{3}$Sc$_{4}$O$_{12}$ are strongly suppressed. In this situation, other types of the magnetic couplings may become important. For example, the absence of the inversion symmetry between nearest-neighbor Cu sites in Ba$_{3}$Cu$_{3}$Sc$_{4}$O$_{12}$ results in the antisymmetric anisotropic exchange interaction that, when taken on its own, will favor orthogonal spin configuration proposed as a candidate ground state of Ba$_{3}$Cu$_{3}$In$_{4}$O$_{12}$~\cite{Volkova}. On the other hand, the non-zero Curie-Weiss temperature $\Theta$ = -70\,K estimated from the high-temperature magnetic susceptibility indicates that isotropic exchange couplings are non-negligible and ferromagnetic. 

Consecutive experimental and theoretical studies~\cite{Dasgupta,Dutton} revealed that intrachain couplings are strong and ferromagnetic, while interchain couplings are weak and antiferromagnetic. However, no conclusive information on the magnetic ground state has been reported, and the relevance of frustrated next-nearest-neighbor couplings within the paper chain remained controversial. Last but not least, anisotropic exchange interactions were disregarded in previous theoretical studies, despite the fact that the orthogonal geometry of CuO$_4$ plaquettes should favor such contributions to the exchange. In the following, we address all these questions in a comprehensive study combining first-principles calculations and accurate numerical simulations of both magnetic ground state and finite-temperature properties. We show that the magnetic ground state of Ba$_{3}$Cu$_{3}$Sc$_{4}$O$_{12}$ is collinear, with ferromagnetic order within the paper chains and antiferromagnetic order between the chains. Magnetic anisotropy determines the direction of the magnetic moment without affecting the ground state qualitatively.

\section{\label{sec:level1}Experimental data}
{\it Structural properties}. 
The polycrystalline sample of Ba$_3$Cu$_3$Sc$_4$O$_{12}$ was prepared at high temperatures via solid state reaction from the stoichiometric mixture of high-purity BaCO$_3$, CuO, and Sc$_2$O$_3$. These reagents were ground, pelletized and fired in alumina crucibles at 850 -- 950 $^{\circ}$C in air for 3 days with regrinding every 24 h. Then the sample was quenched in air to room temperature. Phase purity of the sample was confirmed by powder X-ray diffraction data taken using the ''Radian-2'' diffractometer with CuK$_\alpha$ radiation over a $2\theta$ range of $20-60^{\circ}$. 

The crystal structure of Ba$_{3}$Cu$_{3}$Sc$_{4}$O$_{12}$ presented in Fig.~\ref{im:Crystal} has the tetragonal $I4/mcm$ space group with lattice parameters $a=11.899(2)$\,\AA, $c=8.394(5)$\,\AA, $V=1188(2)$\,\AA$^{3}$, $Z=4$~\cite{Gregory}. The main structural units of the Ba$_{3}$Cu$_{3}$Sc$_{4}$O$_{12}$ compound are CuO$_4$ plaquettes forming a chain along along the $c$ axis. Within an individual chain, there are two nonequivalent crystallographic positions for copper, Cu$^{\rm I}$ and Cu$^{\rm II}$ in the 1:2 ratio.

The copper-oxygen plaquette for Cu$^{\rm I}$ is undistorted and located in the $ab$ plane. The distances between copper and oxygen atoms are 2.028\,\AA. The situation is different for Cu$^{\rm II}$. The plaquette for this copper atom is orthogonal to the $ab$ plane and compressed along the $c$-axis. The copper-oxygen distance for the Cu$^{\rm II}$ atom varies from 1.96\,\AA \, to 2.018\,\AA. The plaquettes in the chain are rotated by $90^{\circ}$ around the $c$ axis with respect to the plaquettes belonging to the neighboring chains. Barium and scandium atoms are located between the chains. As we will show below, scandium orbitals participate in the formation of exchange paths between Cu$^{2+}$ ions in Ba$_{3}$Cu$_{3}$Sc$_{4}$O$_{12}$. Some of the oxygen atoms form bonds between the chains through scandium atoms. These bonds and interactions play significant role in the formation of long-range magnetic order in this system.    

\begin{figure}[!h]
\includegraphics[width=0.5\textwidth]{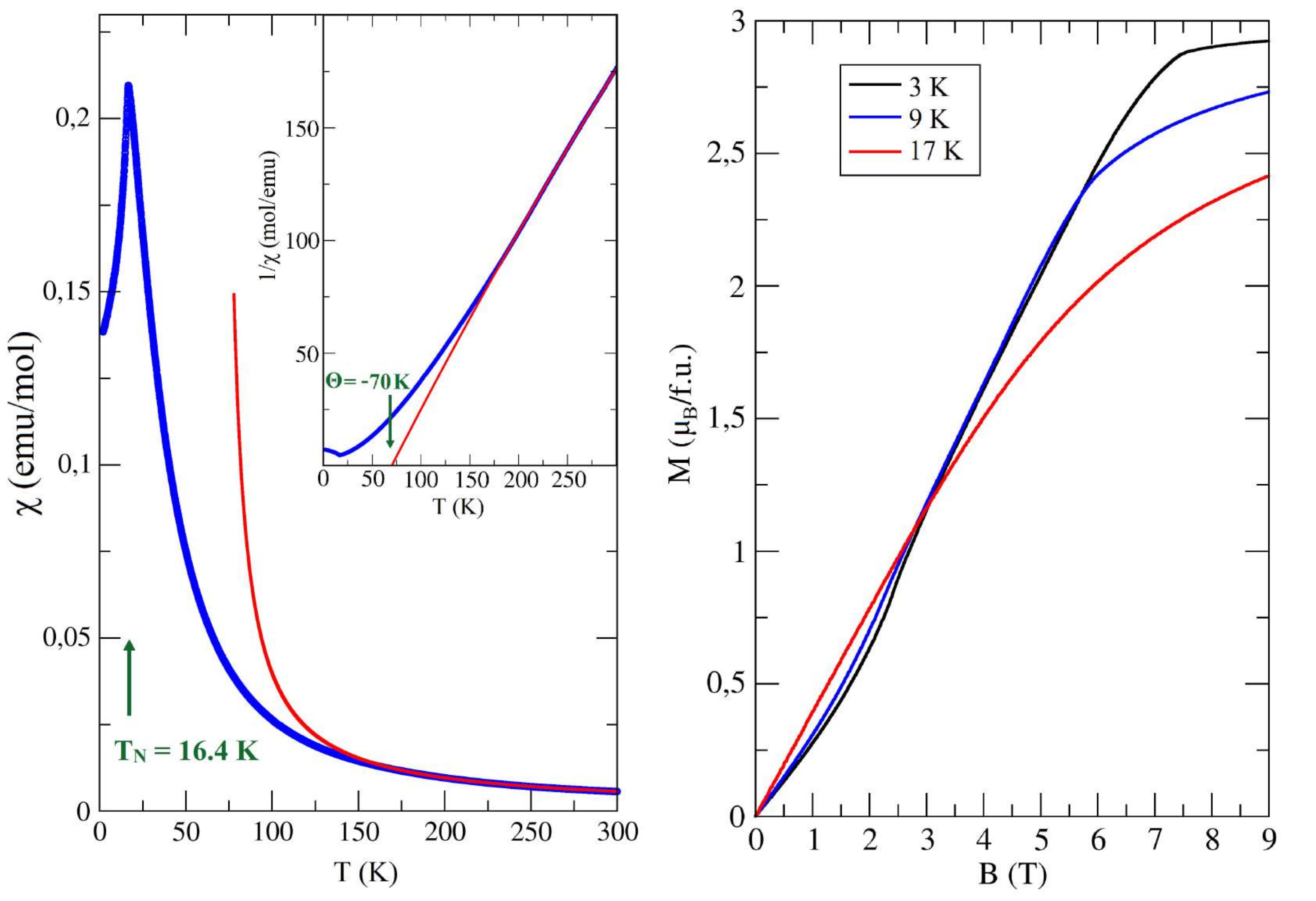}
\caption{(Left panel) Magnetic susceptibility $\chi(T)$ of Ba$_{3}$Cu$_{3}$Sc$_{4}$O$_{12}$ measured in the applied field $B=0.1$\,T. The inset shows the Curie-Weiss approximation with parameters $\Theta = -70\pm2$\,K and $C = (1.18\pm0.03)$\,(emu\,K/mol), as denoted by the red lines. (Right panel) Field dependence of the magnetization measured at different temperatures above and below $T_{N}$. }
\label{im:Thermodynamics}
\end{figure}

{\it Magnetic properties.}
Field and temperature dependencies of the magnetization in Ba$_3$Cu$_3$Sc$_4$O$_{12}$ were measured in the temperature range $2-300$~K and in magnetic fields up to 9~T by means of a Physical Properties Measurement System (PPMS--9T) from Quantum Design. Temperature dependence of the magnetic susceptibility $\chi(T)$ measured in the applied field $B=0.1$\,T is represented in Fig.~\ref{im:Thermodynamics} (left panel). At high temperatures, the susceptibility follows the Curie-Weiss behavior (insert on the left panel of Fig.\ref{im:Thermodynamics}) $\chi=\chi_0+C/(T+\theta)$ with the temperature-independent part $\chi_{0} =(5.5\pm0.6)\times 10^{-4}$~(emu/mol), $\Theta = -70\pm2$\,K and $C=(1.18\pm0.03)$~(emu\,K/mol). From the value of the Curie constant, the $g$-factor can be estimated as 2.05.

The negative value of the Curie-Weiss temperature $\Theta=-70\pm2$\,K suggests that the leading exchange interactions in Ba$_{3}$Cu$_{3}$Sc$_{4}$O$_{12}$ are ferromagnetic, but below 180\,K there is a strong negative deviation from the Curie-Weiss behavior. This deviation indicates the increase of antiferromagnetic correlations, which eventually yield antiferromagnetic long-range order below $T_N = 16.4$\,K. As a consequence, one can observe a sharp peak in the temperature dependence of the magnetic susceptibility $\chi(T)$. 

Field dependence of the magnetization measured at different temperatures is shown in Fig.~\ref{im:Thermodynamics} (right panel). The value of the saturation moment at 3\,K equals to 2.89\,$\mu_B$, which roughly corresponds to three Cu$^{2+}$ ions with the local spin-1/2. Importantly, there is a deviation of the $M(B)$ curve from the linear dependence for the external magnetic fields smaller than 3\,T and at temperatures below $T_N$. The same feature of the magnetization curve was found for the Ba$_{3}$Cu$_{3}$In$_{4}$O$_{12}$ compound~\cite{Volkova}. As we will show below, this peculiarity is due to the inter-atomic symmetric anisotropic exchange interactions.  

\section{\label{sec:level1}Theory}
\subsection{Methods and models}
To describe the electronic structure and magnetic properties of the Ba$_{3}$Cu$_{3}$Sc$_{4}$O$_{12}$ compound, we performed calculations within density functional theory (DFT)~\cite{DFT} using the generalized gradient approximation (GGA) with the PBE exchange-correlation functional~\cite{PBE}. To this end, we employed Quantum Espresso~\cite{Espresso} and Vienna ab initio Simulation Package (VASP)~\cite{Kresse,Furthmuller} codes. In these calculations, we set an energy cutoff in the plane-wave construction to 400\,eV and the energy convergence criteria to 10$^{-4}$\,eV. For the Brillouin-zone integration, a 6$\times$6$\times$6 Monkhorst-Pack mesh was used.

Magnetic behavior of Ba$_{3}$Cu$_{3}$Sc$_{4}$O$_{12}$ is described by the following spin Hamiltonian:
\begin{eqnarray}
\mathcal {\hat{H}} = \sum\limits_{i>j}J_{ij}\hat{\mathbf{S}}_i\hat{\mathbf{S}}_j + \sum\limits_{i>j}\hat{\mathbf{S}}_i\overset{\leftrightarrow}{\Gamma}_{ij}\hat{\mathbf{S}}_j +\sum\limits_{i>j}\mathbf{D}_{ij}[\hat{\mathbf{S}}_i\times\hat{\mathbf{S}}_j] \nonumber \\ 
+\mu_{B}\sum\limits_{i}\hat{\mathbf{S}}_i \tensor g_i\mathbf{B},
\label{eq:Magnetic_model}
\end{eqnarray} 
where $J_{ij}$, $\mathbf{D}_{ij}$ and $\tensor \Gamma_{ij}$ are isotropic, antisymmetric anisotropic and symmetric anisotropic exchange interactions between spins, respectively. $\tensor g_{i}$ is the g-tensor for $i^{\rm th}$ site, and $\mathbf{B}$ is the external magnetic field. 

Exchange interaction parameters can be estimated by different methods. For instance, they can be calculated by using superexchange theory~\cite{Anderson, Aharony,Moriya} on the basis of the DFT results. The corresponding expression for the {\it isotropic exchange interaction} is given by $J_{ij}= 4t^2_{ij}/U$~\cite{Anderson}, where $t_{ij}$ and $U$ are the hopping integral and on-site Coulomb interaction, respectively. This expression describes only the antiferromagnetic kinetic contribution to the total magnetic coupling between the two sites. In the case of cuprates with nearly 90$^{\circ}$ metal-ligand-metal bond angle, there is a ferromagnetic contribution that is formed due to the direct overlap of the neighboring Wannier functions~\cite{Mazurenko, Mila}. All contributions to the isotropic exchange interaction can be accounted for on the level of DFT+$U$ calculations~\cite{Anisimov} by computing differences between total energies of collinear magnetic configurations~\cite{Xiang}. These results will be presented in Section III C.

In order to estimate symmetric ($\tensor \Gamma_{ij}$) and antisymmetric ($\mathbf D_{ij}$) anisotropic interactions, we used the superexchange theory proposed by Moriya~\cite{Moriya} and further developed in Ref.~\onlinecite{Aharony}:
\begin{equation}
\mathbf{D}_{ij} = \frac{i}{U}[{\rm Tr}(\hat {t}_{ij}){\rm Tr}(\hat{t}_{ji}\boldsymbol{\sigma})-{\rm Tr}(\hat{t}_{ji}){\rm Tr}(\hat{t}_{ij}\boldsymbol{\sigma})],
\label{eq:DM-vector}
\end{equation}
\begin{equation}
\overset{\leftrightarrow}{\Gamma}_{ij} = \frac{1}{U}[{\rm Tr}(\hat{t}_{ji}\boldsymbol{\sigma})\otimes {\rm Tr}(\hat{t}_{ij}\boldsymbol{\sigma})-{\rm Tr}(\hat{t}_{ij}\boldsymbol{\sigma})\otimes {\rm Tr}(\hat{t}_{ji}\boldsymbol{\sigma})],
\label{eq:Gamma-matrix}
\end{equation}
where $\boldsymbol{\sigma}$ are Pauli matrices. Within this approach, one calculates the hopping integrals $\hat{t}_{ij}$ taking the spin-orbit coupling into account. In contrast to the previous investigations~\cite{Mila}, where such hoppings were estimated by using the perturbation theory on the spin-orbit coupling, we calculate them from first-principles here, thus treating the spin-orbit coupling on the fully \textit{ab initio} level. To this end, the DFT+SO scheme was used. 

To determine the $g$-tensor in Eq.~\eqref{eq:Magnetic_model}, we use the second-order perturbation theory described in Ref.~\onlinecite{White}. Considering the Wannier function of the $x^2-y^2$ symmetry as the ground-state orbital, one obtains 
\begin{equation}
g^{\mu \nu}_{i} = 2(\delta_{\mu \nu} -\lambda\Lambda^{\mu \nu}_{i}).
\label{eq:g-tensor}
\end{equation}
In this expression, $\lambda\simeq 0.1$\,eV is the spin-orbit coupling constant of the copper atom, and 
\begin{equation}
\Lambda^{\mu \nu}_{i} = \sum\limits_{n} \frac{\langle x^2-y^2|\hat{L}^{\mu}_{i}|n\rangle \langle n|\hat{L}^{\nu}_{i}|x^2-y^2 \rangle}{\epsilon^n_i -\epsilon^{x^2-y^2}_i},
\end{equation}
where $\mu, \nu$ = $x,y,z$.  The index $n$ runs over the excited states of the copper atom, where one of the $d$-orbitals ($xy$, $zx$, $yz$ and $3z^2-r^2$) is half-filled, whereas the $x^2-y^2$ orbital is fully filled. The corresponding orbital energies are defined from DFT calculations.

Due to the complex geometry of Ba$_{3}$Cu$_{3}$Sc$_{4}$O$_{12}$, in our study we use $xyz$ and $abc$ notations for local and global coordinate systems, respectively.

\subsection{\label{sec:level2}DFT results}

\begin{figure}[!b]
\includegraphics[width=0.44\textwidth]{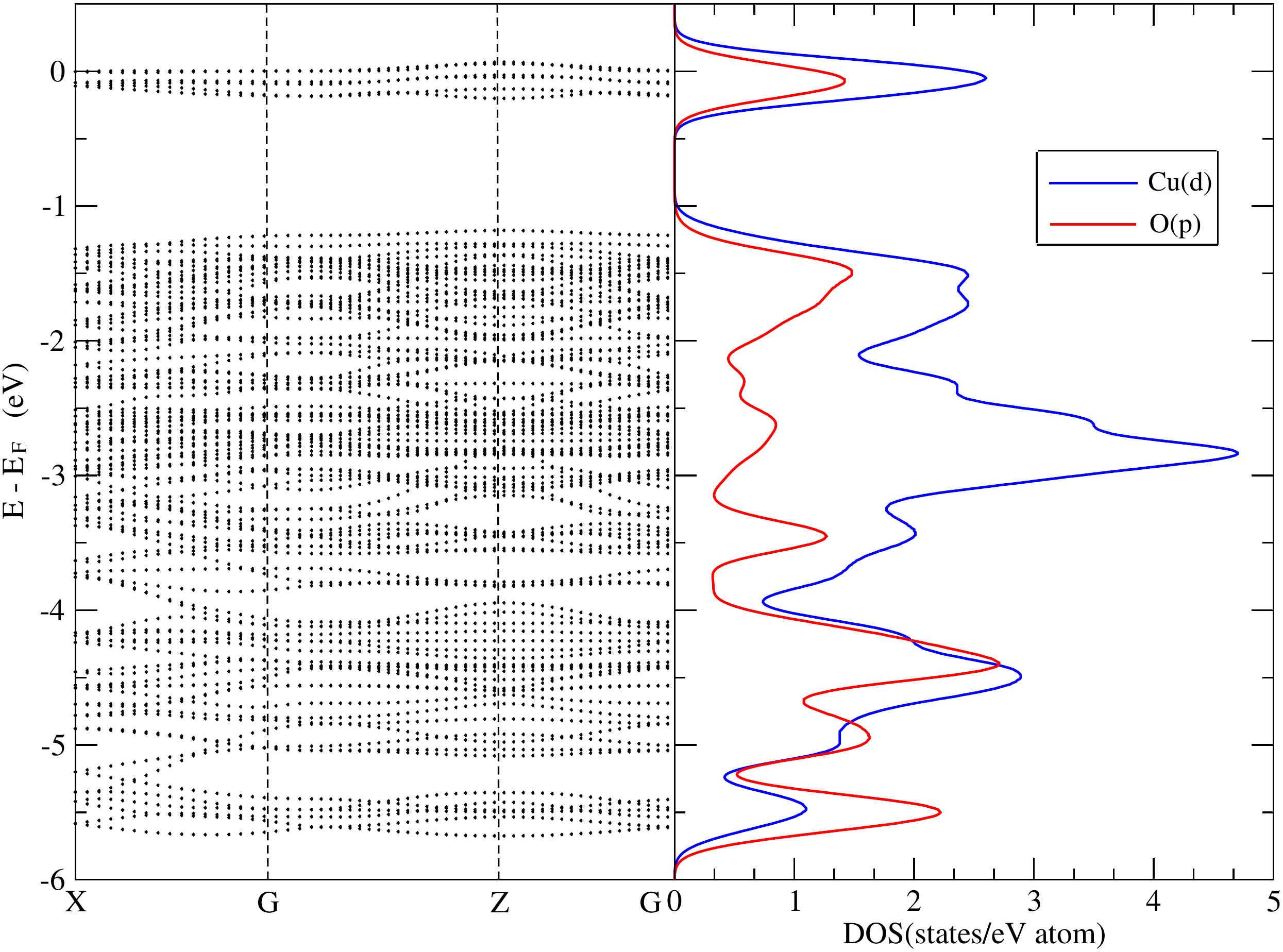}
\caption{(Left) DFT band structure of Ba$_{3}$Cu$_{3}$Sc$_{4}$O$_{12}$. The Fermi level is at zero energy. The high-symmetry $k$-points in the first Brillouin zone are the following: $X=(\frac{1}{2}, 0 ,0)$, $G=(0, 0, 0)$, $Z=(0, 0,-\frac{1}{2})$. (Right) The calculated partial densities of states for copper and oxygen atoms.}
\label{im:Bands}
\end{figure} 

\begin{figure}[!h]
\includegraphics[width=0.44\textwidth]{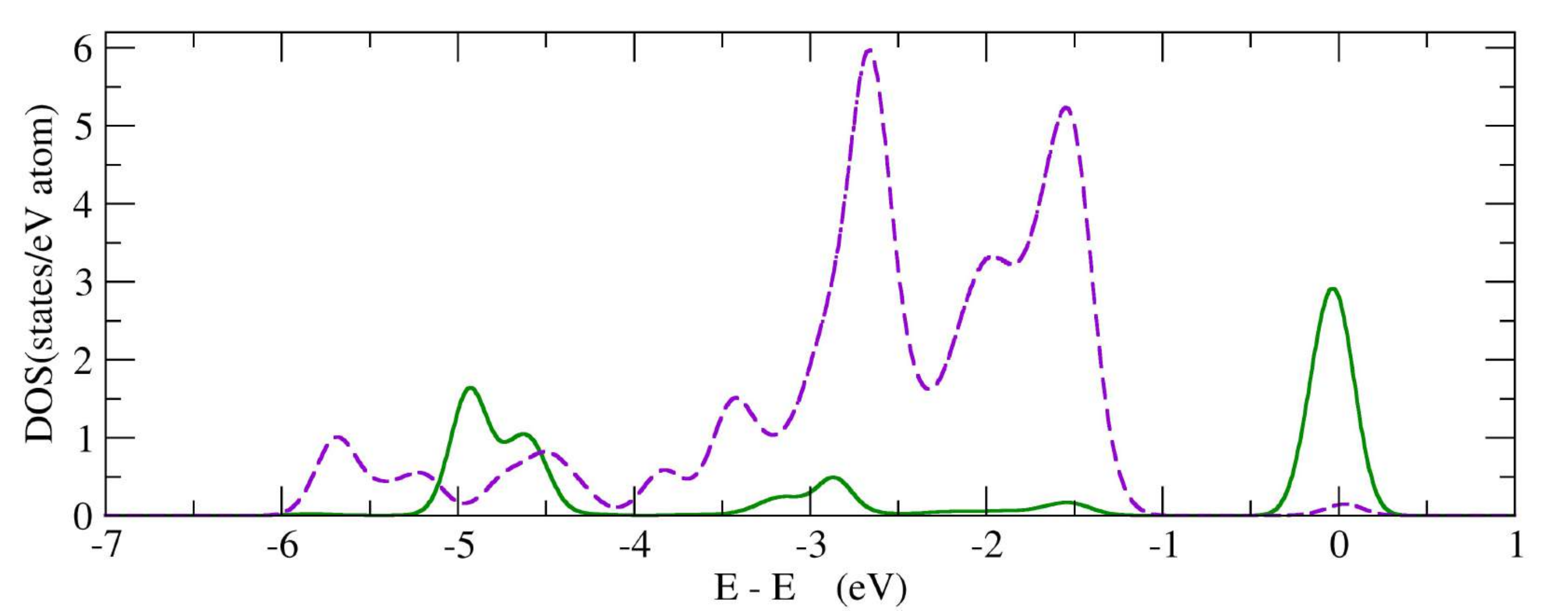}
\caption{The calculated partial densities of states for Cu$^{\rm I}$ atoms. The green solid line corresponds to the $x^2-y^2$ orbitals, the violet dashed line denotes the overall density of states for the four other orbitals ($xy$, $yz$, $xz$, $3z^2-r^2$). The Fermi level is at zero energy. The spectral functions of Cu$^{\rm II}$ in the local coordinate system are close to those calculated for Cu$^{\rm I}$.}
\label{im:Partial_dos}
\end{figure}

{\it Electronic structure.} The calculated GGA electronic structure is presented in Fig.~\ref{im:Bands}. There are six well-separated bands at the Fermi level. According to the calculated partial density of states, these bands correspond to strongly hybridized $3d$ copper and $2p$ oxygen states. Due to a strong overlap of copper orbital of the $x^2-y^2$ symmetry and $2p$ orbitals of the oxygen atoms, the complex molecular orbital is formed. The corresponding bonding and antibonding states are centered at $-4.8$\,eV and 0\,eV, respectively (Fig.~\ref{im:Partial_dos}).

To perform quantitative analysis of the DFT results, we constructed a low-energy model by using the maximally localized Wannier functions (MLWF) procedure, as described in Refs.~\onlinecite{Marzari} and~\onlinecite{Souza}. This procedure is implemented in the Wannier90 code~\cite{Wannier}.

\begin{figure}[!h]
\includegraphics[width=0.44\textwidth]{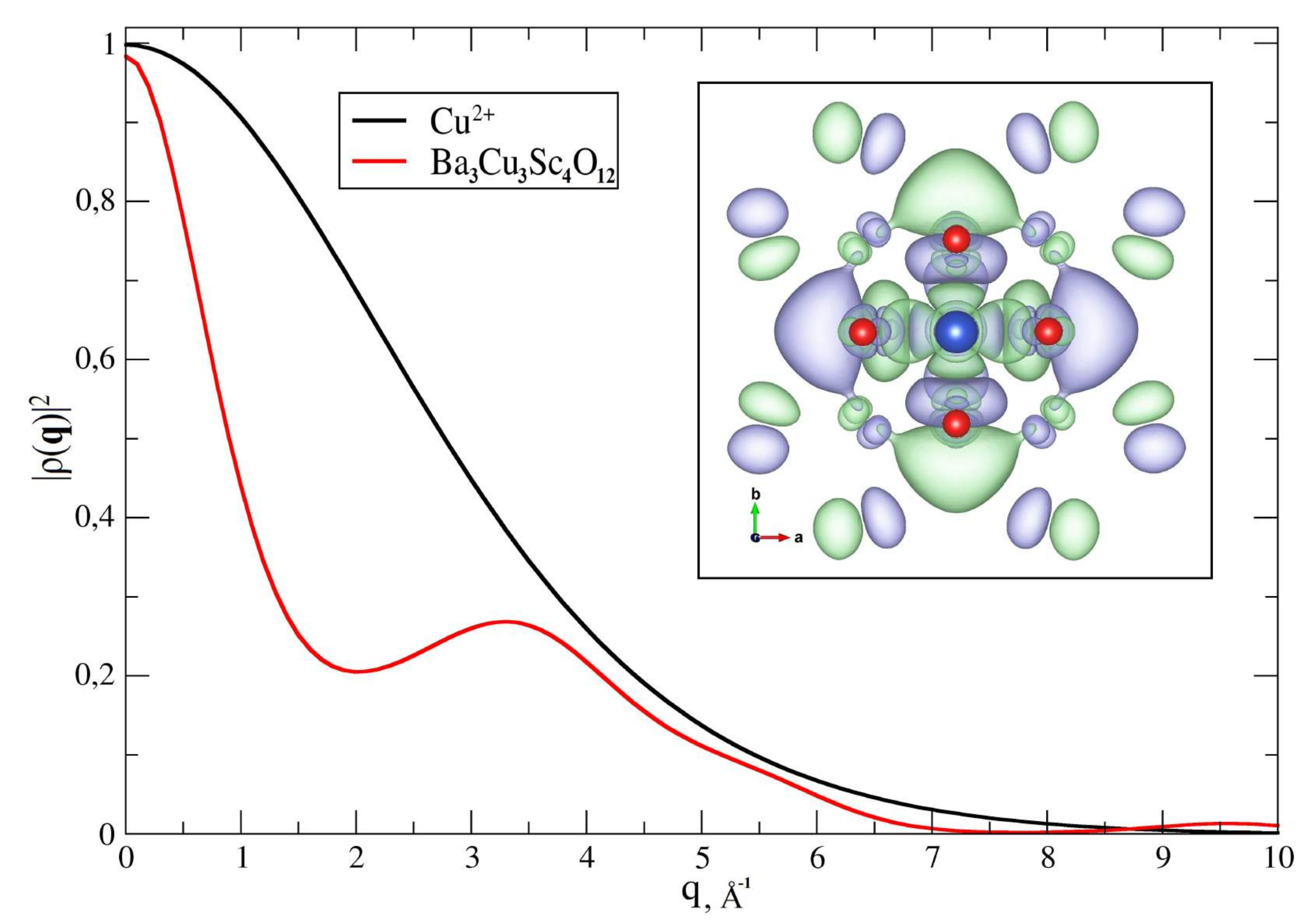}
\caption{Comparison of the ionic Cu$^{2+}$ form factor obtained within the 3-Gaussian approximation~\cite{PJBrown} and covalent form factor calculated for Ba$_{3}$Cu$_{3}$Sc$_{4}$O$_{12}$ by using the constructed Wannier orbitals. The insert shows the Wannier orbital of $x^2-y^2$ symmetry taken from the MLWF procedure, centered on the Cu$^{\rm I}$ atom in the CuO$_4$ plaquette. The blue and red spheres correspond to copper and oxygen atoms, respectively.}
\label{im:F-factor}
\end{figure}

The example of the constructed Wannier orbital is presented in the insert of Fig.~\ref{im:F-factor}. One can see that, being centered at the copper atom, the orbital has large tails at the oxygen sites belonging to the plaquette. The electronic density described by the Wannier function centered on the copper atom can be decomposed as follows: 52\% on the central copper atom, 11\% on each oxygen atom in the plaquette, and approximately 1\% on nearest-neighbor copper atoms. 

Such a delocalization of the electronic density affects the magnetic behavior. For instance, the account of the hybridization with oxygen states changes the magnetic form factor, which is the Fourier transform of the electronic density~\cite{FormFactor}. From Fig.~\ref{im:F-factor} one can see that the covalent form factor decays much faster at small {\bf q} than the pure ionic Cu$^{2+}$ one~\cite{PJBrown}. As we will show below, such a redistribution of the scattering density and the ensuing renormalization of the magnetic form factor plays an important role for simulating neutron scattering spectra.

The hopping matrices between different copper sites in the basis of Wannier functions of $x^2-y^2$ symmetry are represented in Table~\ref{tab:Hoppings}. They agree reasonably well with those obtained by using the tight-binding linearized muffin-tin orbitals method~\cite{Dasgupta}. The difference can be attributed to the fact that the structure of Ba$_{3}$Cu$_{3}$Sc$_{4}$O$_{12}$ is not close-packed. In this case, the TB-LMTO method is very sensitive to computational details, such as the radii of muffin-tin spheres and the composition of the valence shell. We demonstrate this in the Supplementary Material.  

 \begin{table}[!b]
\centering
\caption [Bset]{Comparison of the hopping integrals (in\,meV) calculated by using the MLWF method in this work and the downfolding procedure described in Ref.~\onlinecite{Dasgupta}. The interaction paths are labeled in accordance with Fig.~\ref{im:Exchange} (left) and additionally denoted by Cu--Cu distances $d_{\rm Cu-Cu}$ (in\,\AA).}
\label {basisset}
\begin{ruledtabular}
\begin {tabular}{llcc}
  path & d$_{\rm Cu-Cu}$ & MLWF  &  Ref.\onlinecite{Dasgupta}  \\
 \hline
1&  2.74  &5.45         &16.3\\
2&  3.53  &38.80        &53.0\\
3&  4.19  &40.07        &45.0 \\
4&  4.88  &15.84        &13.6 \\
5&  6.88  &9.68         &8.1\\
6&  6.97  &6.25         &5.4\\
7a& 8.39  &7.36         &8.0\\
7b& 8.39  &13.76        &13.6\\
8&  8.41  &18.23        &19.0\\
9&  9.15  &14.21        &14.9\\
\end {tabular}
\end{ruledtabular}
\label{tab:Hoppings}
\end {table}
The hopping integral $t_{1}$, which describes the interaction between the nearest-neighbor copper atoms within the chain, is considerably smaller than other interactions. It is due to the cosine-type dependence of the hopping integral on the Cu--O--Cu bond angle, $t \sim \cos \alpha$. In Ba$_{3}$Cu$_{3}$Sc$_{4}$O$_{12}$, $\alpha$ is about 86.8$^\circ$. As a result, the corresponding kinetic exchange interaction is strongly suppressed for the nearly orthogonal superexchange path. According to the obtained hopping integrals, the strongest interactions are $t_2$ and $t_3$ between Wannier functions belonging to parallel CuO$_4$ planes. Among the inter-chain couplings, the largest one is between the Cu$^{\rm I}$ atoms, $t_{8}$. As we will show below, it is mainly responsible for three-dimensional magnetic ordering in Ba$_{3}$Cu$_{3}$Sc$_{4}$O$_{12}$. 

{\it Magnetic interactions.}
With the hopping parameters described above, one can calculate isotropic exchange interactions using Anderson's formula for superexchange interaction $J_{ij}= 4t{^2}_{ij}/U$~\cite{Anderson}, where $U$ is an effective on-site Coulomb interaction for the one-band model. These results are represented in Table~\ref{tab:Isotropic_exchange}. The $U$ value of 4\,eV was estimated within constrained calculations by using TB-LMTO-ASA~\cite{LMTO,ASA} described in the Supplementary Material. 
\begin{figure}[!h]
\includegraphics[width=0.44\textwidth]{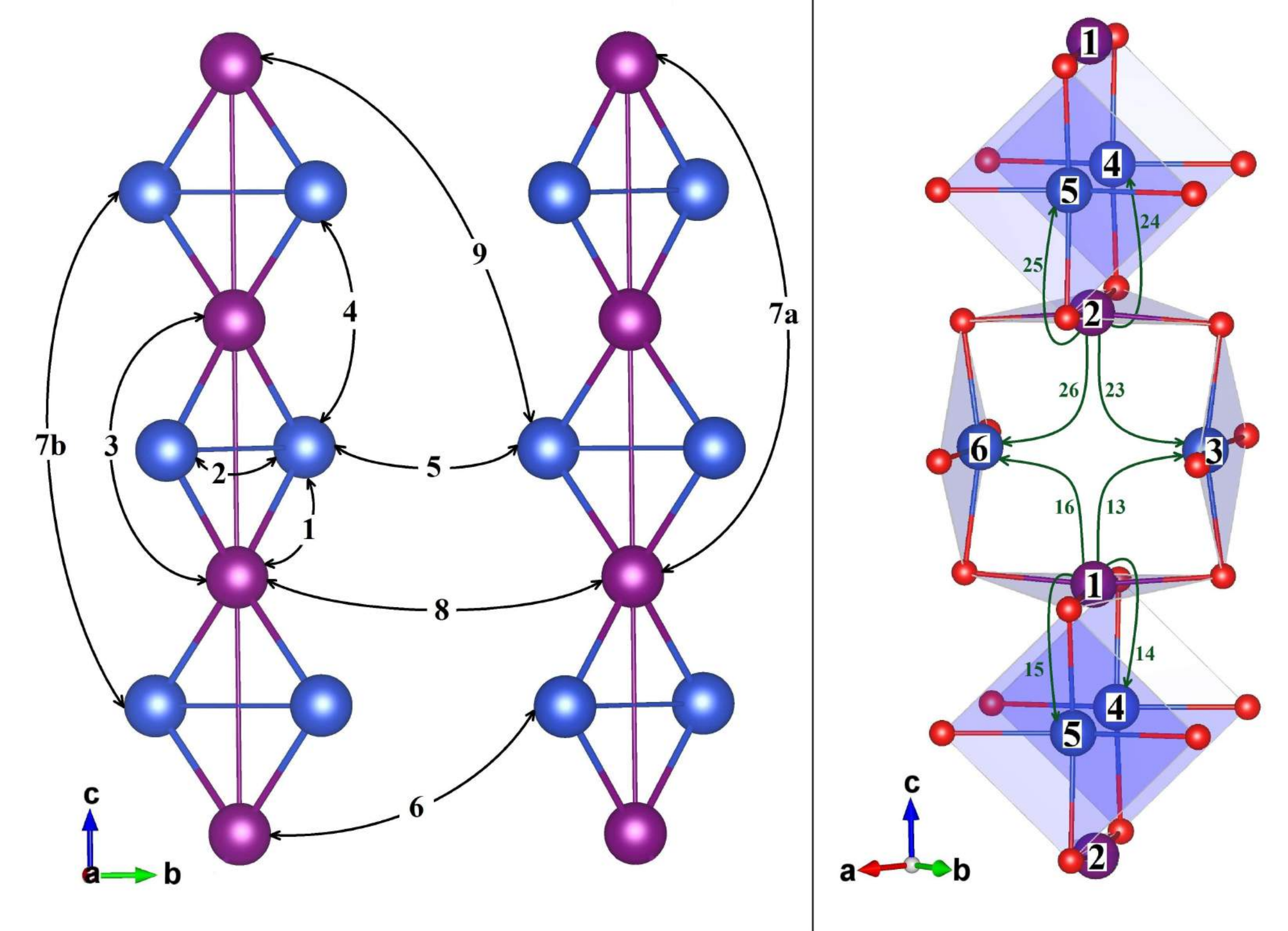}
\caption{(Left panel) The paths of Cu--Cu interactions. (Right panel) Schematic representation of anisotropic exchange interactions within the chain. }
\label{im:Exchange}
\end{figure} 

One can see that the kinetic antiferromagnetic exchange interaction between nearest neighbors in the chain is negligibly small ($\sim$ 0.01 meV). At the same time, there is a strong ferromagnetic contribution due to the direct overlap of the neighboring Wannier functions on the oxygen atoms. Starting with the general Hubbard model containing different types of the inter-site Coulomb interactions~\cite{Mazurenko}, one can show that the ferromagnetic contributions are of the form of the direct exchange interaction between the Wannier functions. For instance, the oxygen contribution to the total exchange interaction is expressed as $J^{\rm FM}_{ij} = -2 \beta^4 J_{H}^{p}$, where $\beta$ is the contribution of oxygen states to the magnetic Wannier function, and $J_{H}^{p}$ is the intra-atomic exchange interaction for the oxygen atom. Since the resulting exchange interaction is sensitive to the structure of the Wannier function, at the moment we restrict ourselves to estimating the order of magnitude of $J^{\rm FM}_{ij}$, which is a few tens of meV. Given the small size of the antiferromagnetic contribution, we conclude that the interaction between nearest neighbors is ferromagnetic.    

\begin{figure}[!h]
\includegraphics[width=0.45\textwidth]{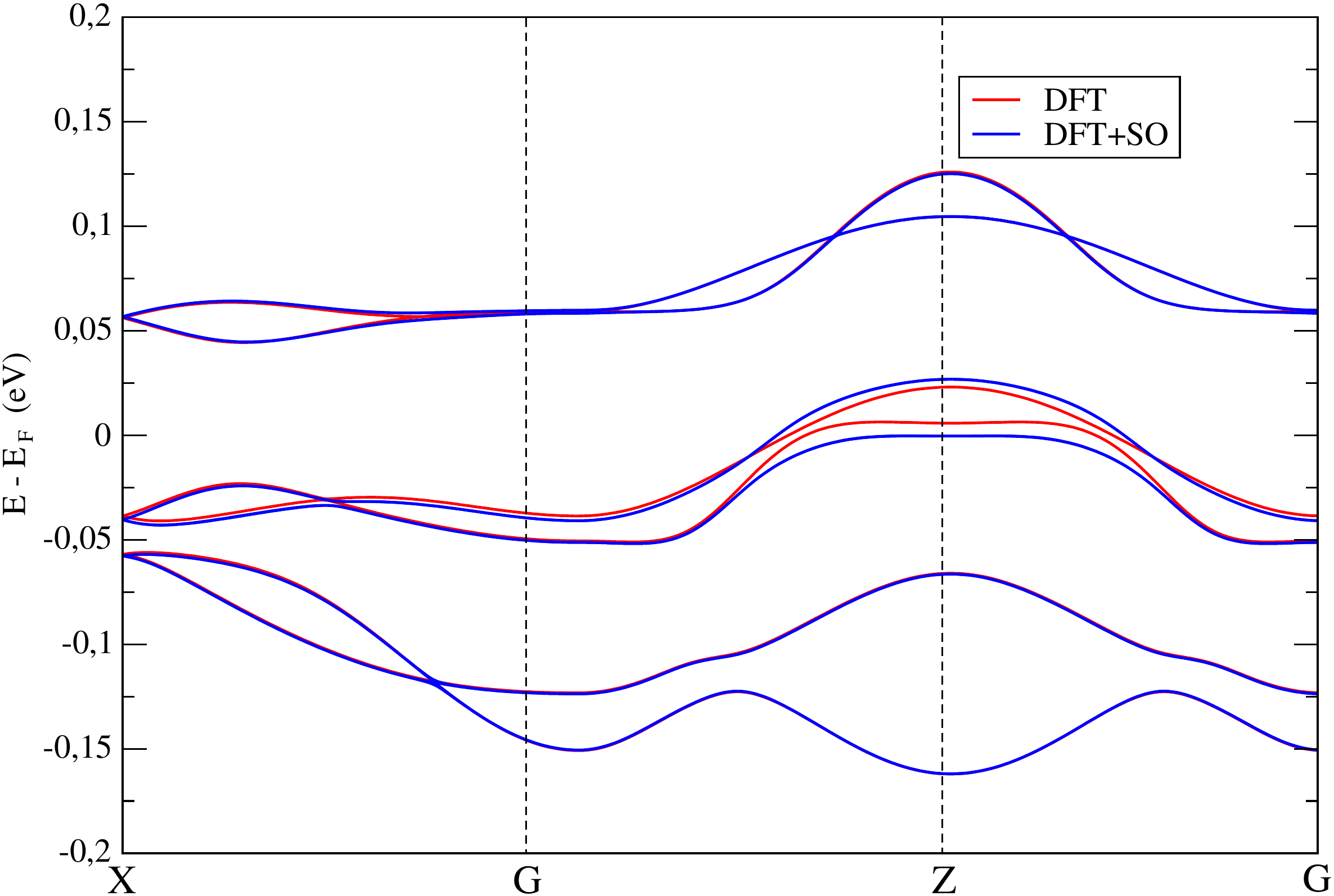}
\caption{Comparison of the band structures for Ba$_{3}$Cu$_{3}$Sc$_{4}$O$_{12}$ calculated by DFT and DFT+SO methods. }
\label{im:Spin-orbit}
\end{figure}

In order to estimate the anisotropic exchange interactions, the DFT+SO scheme was used. The resulting band structure is presented in Fig.~\ref{im:Spin-orbit}. One can see that the account of the spin-orbit coupling leads to weak band splittings at the $Z$ point.  
In contrast to the DFT results, the hopping matrix between the $x^2-y^2$ Wannier functions calculated within the DFT+SO approach has non-diagonal complex parts that are triggered by the mixing of excited states by the spin-orbit coupling. For instance, in the case of the 1-3 bond we have (in meV):
\[ \hat t_{13}= \left( \begin{array}{cc} 
 6.1 & 3.7 + 3.7 i   \\
-3.7 + 3.7 i & 6.1       
\end{array} \right). 
\]
Namely, the non-zero $t^{\uparrow \downarrow}_{13}$ and $t^{\downarrow \uparrow}_{13}$ components are responsible for the anisotropic couplings for this bond.  

In the global coordinate system $abc$, the symmetric anisotropic tensors calculated by using Eq.~\eqref{eq:Gamma-matrix} for the nearest-neighbors bonds are given by (in meV):

\[\tensor \Gamma_{15}= \tensor \Gamma_{24}= \tensor \Gamma_{14}= \tensor \Gamma_{25}= \left( \begin{array}{ccc} 
 0.009 & -0.028 & 0  \\
-0.028 & 0.009 & 0   \\
 0 & 0 & -0.018   
\end{array} \right), 
\]

\[\tensor \Gamma_{13}= \tensor \Gamma_{26}= \tensor \Gamma_{16}= \tensor \Gamma_{23}= \left( \begin{array}{ccc} 
 0.009 & 0.028 & 0  \\
 0.028 & 0.009 & 0  \\
 0 & 0 & -0.018   
\end{array} \right). 
\]
Our calculations revealed that the anisotropic $\tensor \Gamma_{ij}$ for other pairs of spins are negligibly small and can be excluded from further consideration. 

We first analyze symmetric anisotropic interactions between Cu$^{\rm I}$ and its magnetic environment. The calculated non-diagonal elements of $\tensor \Gamma_{1j}$ in the global coordinate system are bond-dependent and correspond to a quantum compass model~\cite{Sr2IrO4}. The clock-wise rotation of the coordinate system by 45$^{\circ}$ around the $c$ axis leads to the diagonal form of the anisotropic tensors, $\Gamma^{a'a'}_{1j} = \Gamma^{aa}_{1j} \pm |\Gamma^{ab}_{1j}|$ and $\Gamma^{b'b'}_{1j} = \Gamma^{bb}_{1j} \mp |\Gamma^{ab}_{1j}|$, where upper (lower) sign stands for the bonds 1--4 and 1--5 (1--3 and 1--6). One finds that $\Gamma^{a'a'}_{1j}$ = $\Gamma^{c'c'}_{1j}$ ($\Gamma^{b'b'}_{1j}$ = $\Gamma^{c'c'}_{1j}$) for $j$=3,6 (for $j$=4,5). Given the ferromagnetic nature of the coupling $J_1$, the lowest component of the anisotropy defines the magnetic easy axis. By calculating the sum of anisotropic exchange couplings $\sum_{j} \Gamma_{1j}$, we find that $c$-axis is the magnetic easy axis for Cu$^{\rm I}$. 

The anisotropic couplings between the Cu$^{\rm II}$ ion and its nearest neighbors are different. We obtain easy-plane anisotropy of $a'c'$ and $b'c'$ symmetry for atoms 3(6) and 4(5), respectively. Based on the analysis of the calculated symmetric anisotropic exchange interactions, we conclude that the $c$-axis is the easy axis for the total magnetization of the system.

In turn, Dzyaloshinskii-Moriya vectors were calculated by using Eq.~\eqref{eq:DM-vector}. Similar to $\tensor \Gamma$, we found that the strongest DM couplings are between nearest neighbors within the chain, $\mathbf{D}_{24}=\mathbf{D}_{15}=d(-1,1,0)$, $\mathbf{D}_{25}=\mathbf{D}_{14}=d(1,-1,0)$, $\mathbf{D}_{13}=\mathbf{D}_{26}=d(1,1,0)$, and $\mathbf{D}_{16}=\mathbf{D}_{23}=d(-1,-1,0)$, where $d$ = 0.046 meV. 
Nearest-neighbor Dzyaloshinskii-Moriya interactions within the chain will typically produce either spiral or canted configuration for uniform and staggered arrangement of the DM vectors, respectively. In the case of Ba$_{3}$Cu$_{3}$Sc$_{4}$O$_{12}$, the Dzyaloshinskii-Moriya interactions along the 2--6--1 path tend to misalign the ferromagnetic order of spins 1 and 2. However, this effect is fully compensated by the Dzyaloshinskii-Moriya interactions of different sign along the 2--3--1 bond that will produce same misalignment but in the opposite direction. Thus the antisymmetric anisotropic exchange interactions do not distort the collinear ferromagnetic configuration along the chain.     

The next important quantity that we analyze is the $g$-tensor. Experimentally, it was measured for the Ba$_{3}$Cu$_{3}$In$_{4}$O$_{12}$ compound~\cite{Volkova}. It was shown that the $g$-factor value for the direction parallel to the normal to the CuO$_{4}$ plaquettes is larger than in the perpendicular direction, 2.15 and 2.08, respectively. Since the uniaxial symmetry axes of three CuO$_{4}$ plaquettes in the unit cell are mutually orthogonal, the corresponding $g$-tensors should be mutually orthogonal as well. 

We have calculated the components of the $g$-tensor by using perturbation theory on the spin-orbit coupling, Eq.~\eqref{eq:g-tensor}. To this end, the energies of the Wannier orbitals for copper atoms presented in Table~\ref{tab:Wannier_energies} were derived from the DFT calculations. For Cu$^{\rm II}$ the splitting between the $x^2-y^2$ and $xy$ orbitals is 0.5\,eV larger than that for Cu$^{\rm I}$. This difference can be explained by the difference in the copper-oxygen distances within the plaquette. In the case of the Cu$^{\rm II}$ plaquette, the shorter Cu--O bond leads to a stronger hybridization between the Cu-$x^2-y^2$ and O-$2p$ states. 

\begin{table}[!h]
\centering
\caption [Bset]{ The calculated energies $\epsilon_n$ (in eV) of the Wannier orbitals centered at copper atoms in the local coordinate frame. The results were obtained by  
using maximally localized Wannier functions procedure. The atom numeration is taken from Fig.\ref{im:Exchange} (right panel).}
\label {basisset}
\begin{ruledtabular}
\begin {tabular}{lccccc}
Atom & $3z^2 - r^2$ & $xz$ & $yz$ & $x^2-y^2$ & $xy$ \\
\hline
Cu$^{\rm I}$ & $-2.51$ & $-2.47$ & $-2.47$ & 0 & $-2.70$ \\
Cu$^{\rm II}$ & $-2.81$ & $-2.97$ & $-2.84$ & 0 & $-3.22$ \\
\end {tabular}
\end{ruledtabular}
\label{tab:Wannier_energies}
\end {table}

The resulting $g$-tensors in the global coordinate frame are as follows (the atoms are numbered according to the right panel of Fig.\ref{im:Exchange}):

\[\tensor g_{1,2} = \left( \begin{array}{ccc} 
 2.082 & 0 & 0  \\
 0 & 2.082 & 0  \\
 0 & 0 & 2.296   
\end{array} \right), 
\]

\[\tensor g_{3,6} = \left( \begin{array}{ccc} 
 2.158 & -0.09 & 0  \\
 -0.09 & 2.158 & 0  \\
 0 & 0 & 2.070   
\end{array} \right), 
\]

\[\tensor g_{4,5} = \left( \begin{array}{ccc} 
 2.158 & 0.090 & 0  \\
 0.090 & 2.158 & 0  \\
 0 & 0 & 2.070   
\end{array} \right). 
\]

The components of the $g$-tensor obey the symmetry of the system. From $g$-tensors for each copper atom one can estimate the average values for the whole system, arriving at $g_{c}$ = 2.132 and $g_{ab}$ = 2.145. These values will be used in quantum Monte Carlo (QMC) simulations, which are described in the next section. 

\subsection{\label{sec:level3}DFT+$U$ results}
The analysis of the magnetic interactions presented in the previous section is based on the perturbation theory $t \ll U$. Thus the next step of our investigation was to take into account kinetic processes and Coulomb correlations on equal footing within the DFT+$U$ approach. Such calculations revealed that the Ba$_{3}$Cu$_{3}$Sc$_{4}$O$_{12}$ compound is insulating with the band gap of about 2\,eV. The obtained value of the copper magnetic moment is about 0.65\,$\mu_{B}$, which is much smaller than 1\,$\mu_{B}$ expected for $S = \frac{1}{2}$. It is the result of the strong hybridization between the Cu-$3d$ and O-$2p$ states.  

On the basis of the DFT+$U$ calculations we calculated exchange interactions as differences between total energies of collinear spin configurations. Table \ref{tab:Isotropic_exchange} lists leading isotropic exchange interactions for different values of the on-site Coulomb interaction $\tilde U$ in DFT+$U$. In accordance with DFT results, the largest interaction, $J_1$, is ferromagnetic. Among the inter-chain interactions we are mainly interested in $J_{8}$ that runs between Cu$^{\rm I}$ atoms belonging to the plaquettes, which are perpendicular to the chain axis. In Fig.~\ref{im:Crystal}, one can see that the corresponding magnetic $x^2-y^2$ orbitals from neighboring chains point toward each other, which guarantees the largest overlap. 

\begin{table}[!h]
\centering
\caption [Bset]{The leading isotropic exchange interactions (in meV) in Ba$_{3}$Cu$_{3}$Sc$_{4}$O$_{12}$ calculated by using total energies difference method on the basis of the DFT+$U$ results and estimated by means of the Anderson's superexchange theory. The notation of the exchange interactions is taken from Fig.\ref{im:Exchange} (left).   }
\label {basisset}
\begin{ruledtabular}
\begin{tabular}{lcrrrr}
 N & $ 4t^2/U$  & $\tilde U$=8 eV  & $\tilde U$=10 eV & $\tilde U$=11 eV & Ref.\onlinecite{Dasgupta} ($\tilde U$=8 eV)  \\
  \hline
 $J_1$ &  0.03   & $-15.71$ & $-11.98$ & $-9.88$ & $-13.88$ \\
 $J_2$ &  1.51   & 1.95   & 1.31  & 1.06  &  6.93  \\ 
 $J_3$ &  1.61   & 0.91   & 0.51  & 0.35  &  2.74  \\
 $J_4$ &  0.25   &$-0.15$ &$-0.16$&$-0.15$&  2.53  \\
 $J_8$ &  0.33   & 0.59   & 0.36  & 0.21  &  3.89  \\
\end{tabular}
\end{ruledtabular}
\label{tab:Isotropic_exchange}
\end {table}

According to Anderson's superexchange theory, the distant intra- and inter-chain interactions $J_{4}$ and $J_{8}$ are much smaller than $J_{2}$ (Table~\ref{tab:Isotropic_exchange}). This scenario is well reproduced by our DFT+$U$ calculations, in contrast to the DFT+$U$ results reported in Ref.~\onlinecite{Dasgupta}. Moreover, we observe a large difference between the values of antiferromagnetic exchange interactions $J_{2}$, $J_{3}$, $J_{4}$ and $J_{8}$.

The Curie-Weiss temperature was estimated by using isotropic exchange interactions, $\Theta_{\rm Cu^{\rm I}}=\frac{1}{4k_B}(4J_1+2J_3+4J_8), \Theta_{\rm Cu^{\rm II}}= \frac{1}{4k_B}(2J_1+J_2+2J_4)$, where $k_B$ is the Boltzmann constant. Taking into account the number of atoms of each type in the unit cell, the averaged Curie-Weiss temperature can be defined as $\Theta= \frac{1}{3}(\Theta_{\rm Cu^{\rm I}}+2\Theta_{\rm Cu^{\rm II}})$. For $\tilde U$=11\,eV we obtain $\Theta=-73$\,K, which is consistent with the experimental value of $-70\pm 2$\,K. 

\subsection{\label{sec:level4}DFT+$U$+SO results}
The account of the spin-orbit coupling within the DFT+$U$ scheme gives us an opportunity to study orbital magnetism in Ba$_{3}$Cu$_{3}$Sc$_{4}$O$_{12}$. The size of the orbital magnetic moment on the given copper site in the chain depends on the direction of the total magnetization of the system. For instance, we get $L^z_{\rm Cu^{\rm I}}$ = 0.175~$\mu_B$ and $L^z_{\rm Cu^{\rm II}}$ = 0.045~$\mu_B$ when the total magnetic moment is along the $c$-axis. Thus, the preferable direction for the orbital moment is normal to the CuO$_4$ plaquette. This effect can be explained on the level of the crystal-field splitting revealed by the DFT calculations (Table~\ref{tab:Wannier_energies}). The energies of the $xy$ and $yz$ ($xz$) orbitals are close to each other. However, the corresponding matrix elements of the orbital momentum operator are very different, $\langle xy | L^{z}_{i} | x^2-y^2 \rangle$ = $2i$ and $\langle x^2-y^2 | L^{y}_{i} | xz \rangle$ = $i$. Therefore, within the perturbation theory on the spin-orbit coupling, the preferred positioning of the orbital moment along the normal of the plaquette is confirmed.    

From DFT+$U$+SO calculations we also estimated diagonal components of the anisotropic $\tensor \Gamma$ by using the total energies difference method. The obtained   diagonal components are $\Gamma^{aa}_{13} = \Gamma^{bb}_{13} = 0.05$\,meV and $\Gamma^{cc}_{13} = -0.1$\,meV. The symmetry of the resulting $\tensor \Gamma$ agrees with those estimated by using the perturbation theory Eq.~(\ref{eq:Gamma-matrix}). However, the absolute values of the $\tensor \Gamma$ elements obtained in DFT+$U$+SO are about five times larger than those from DFT+SO calculations presented in the previous section. Such a difference can be naturally explained by the fact that the perturbation theory results were obtained for the one-band model with spin-orbit coupling. The account of the complex multi-orbital structure of the copper atoms within DFT+$U$+SO scheme gives additional contributions to the inter-site anisotropic exchange interaction. These contributions originate from intra-atomic Hund's rule exchange, as discussed in Ref.~\onlinecite{Aharony,Sr2IrO4}. 

\section{\label{sec:level1}Comparison with experiment}
In this section, we investigate the magnetic model of Ba$_{3}$Cu$_{3}$Sc$_{4}$O$_{12}$, define the ground-state spin configuration, and describe magnetic properties observed experimentally. 

\begin{figure*}[]
\includegraphics[width=0.30\textwidth]{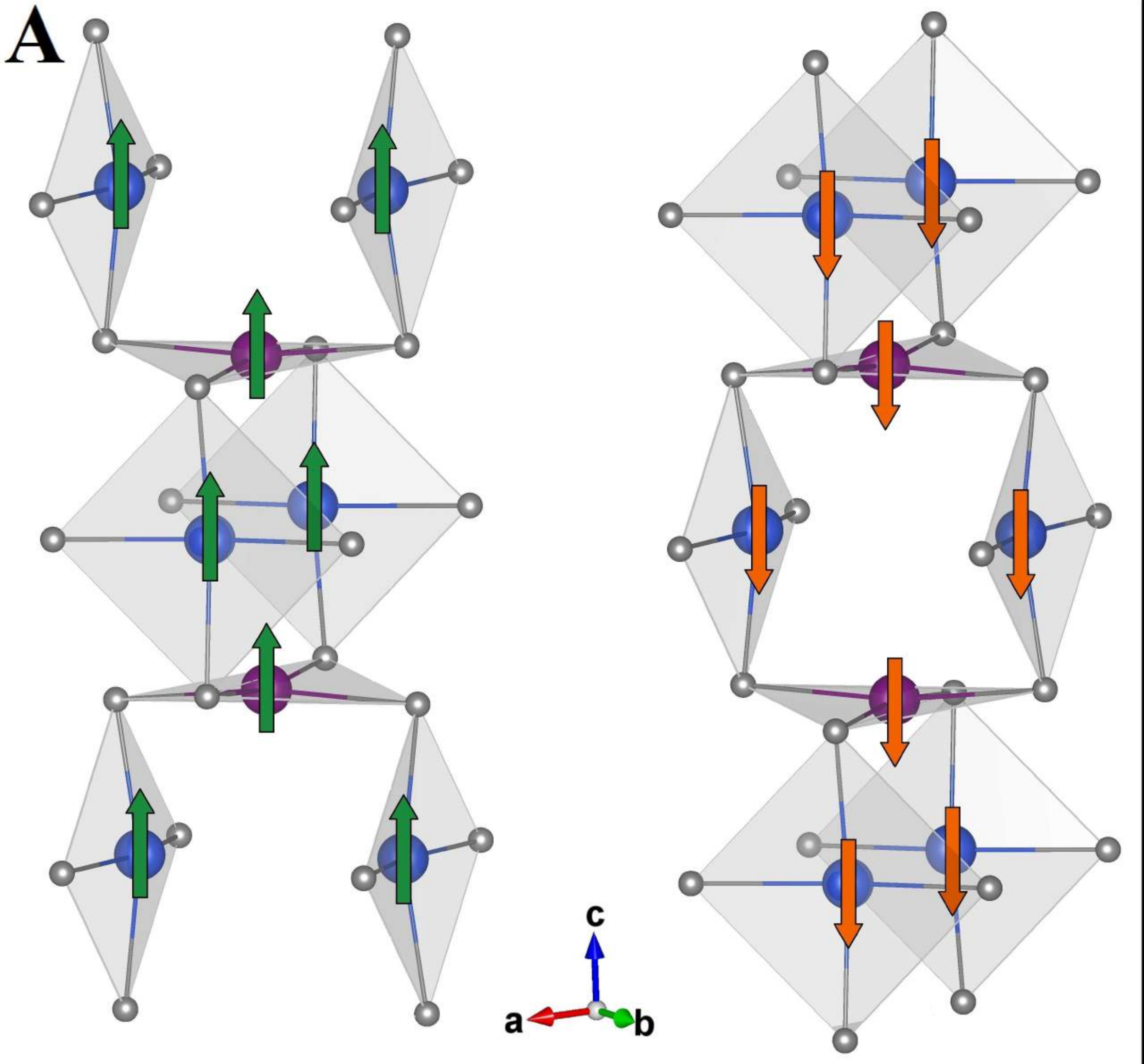}
\includegraphics[width=0.30\textwidth]{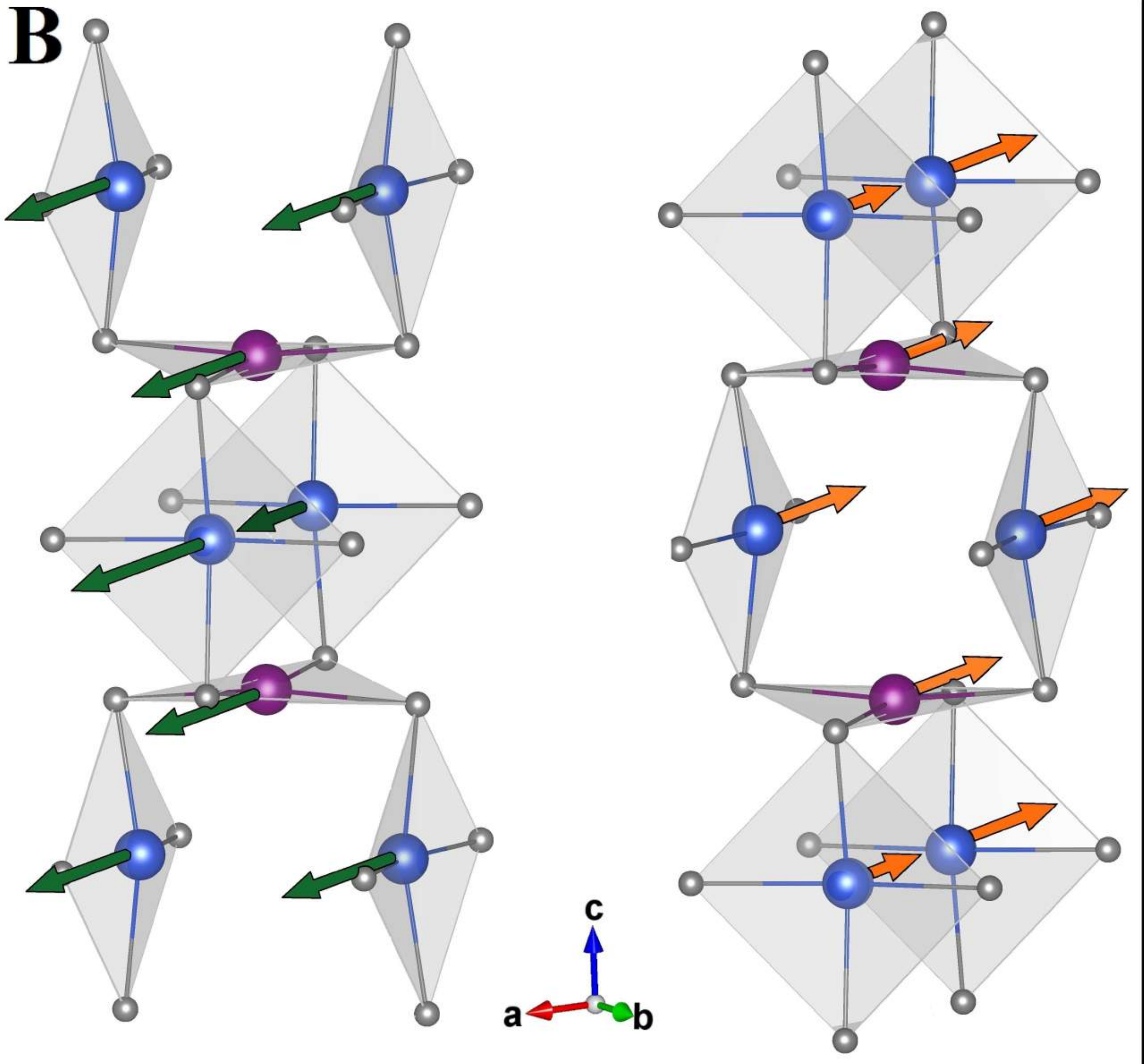}
\includegraphics[width=0.30\textwidth]{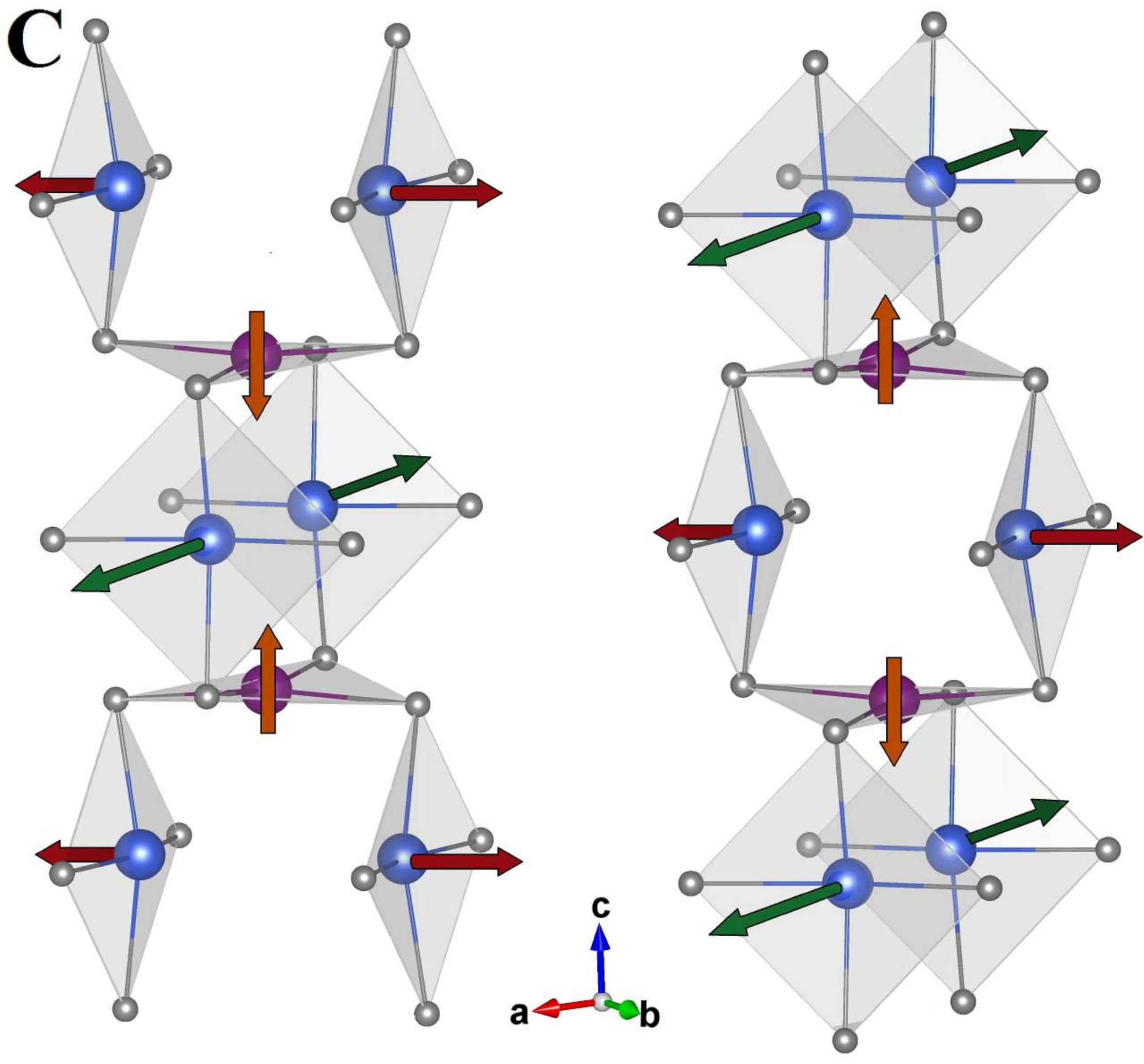}
\caption{Three candidate ground-state magnetic configurations of Ba$_{3}$Cu$_{3}$Sc$_{4}$O$_{12}$ considered in this work.}
\label{im:Local_minimum}
\end{figure*}

\subsection{\label{sec:level4}Classical ground state}
We start with the solution of the magnetic model considering spins as classical vectors. Three magnetic configurations presented in Fig.~\ref{im:Local_minimum} are considered:
\begin{itemize}
 \item $\mathbf{A}$ -- ferromagnetic chains with spins directed along the $c$-axis  (Fig.~\ref{im:Local_minimum}A);
 \item $\mathbf{B}$ -- ferromagnetic chains with spins directed along the $a + b$ axis (Fig.~\ref{im:Local_minimum}B);
 \item $\mathbf{C}$ -- orthogonal spin arrangement (Fig.\ref{im:Local_minimum}C) proposed in Ref.~\onlinecite{Volkova}.
\end{itemize}
In all three cases, antiferromagnetic ordering between the chains is assumed.

We found that the configuration $\mathbf{A}$ has the lowest energy. The energy of the configuration $\mathbf{B}$ is 0.028\,meV higher according to the the difference $\Gamma^{aa}_{13} - \Gamma^{cc}_{13}$. The configuration $\mathbf{C}$ has the highest energy, because the leading exchange interaction $J_{1}$ does not contribute to the total energy of this orthogonal spin configuration. 

\subsection{\label{sec:level2}Quantum Monte Carlo simulations}
Since the absolute value of $J_{1}$ is much larger than $J_{2}$ and $J_{3}$, we can tentatively neglect frustration by second-neighbor interactions within the chain and proceed to the numerical treatment of the quantum spin model by quantum Monte Carlo (QMC) simulations. 

Thermodynamic properties were obtained using stochastic series expansion (SSE)°\cite{sse} method implemented in the loop\cite{loop} algorithm of the ALPS°\cite{ALPS} simulation package.
\begin{figure}[!h]
\includegraphics[width=0.44\textwidth]{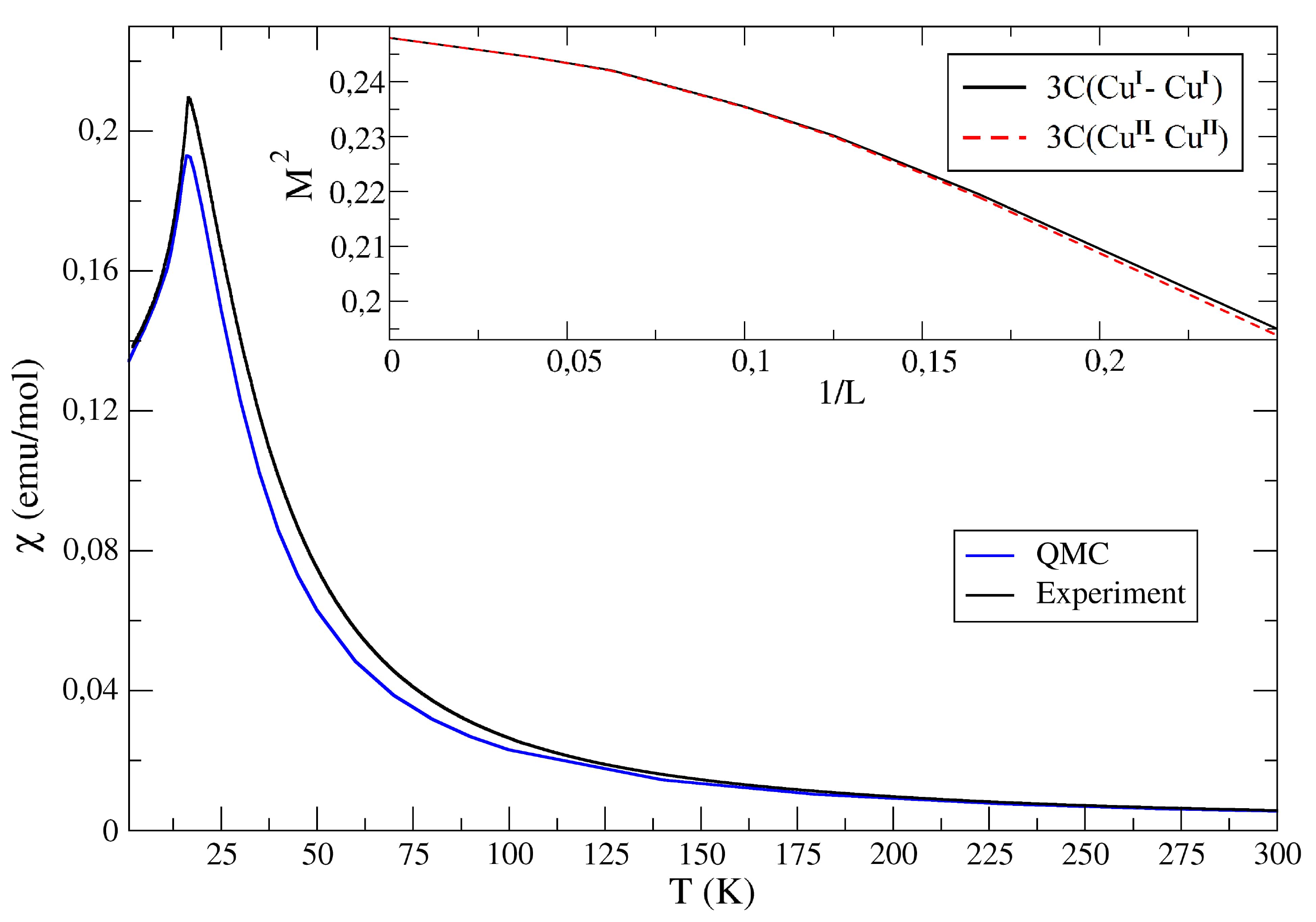}
\caption{Results of QMC simulation for the magnetic susceptibility in comparison with experiment. The insert shows spin-spin correlation functions depending on the inverse linear size of the system $L$.}
\label{im:Susceptibility}
\end{figure}
We performed simulations for $L/2\times L/2\times L$ finite lattices with $L$ $\le$ 20 and periodic boundary conditions. In these calculations, we take into account the leading intra-chain interaction $J_{1}$ and inter-chain couplings $J_{8}$ estimated from DFT+$U$. To take other inter-chain couplings into account, an effective interaction of the $J_{5}$ type between Cu$^{\rm II}$ atoms is introduced into our model. The value of this interaction was estimated by using superexchange theory,
 $J_5 =  4t_5^2/U$ = 0.093\,meV. To include the effects of the spin-orbit coupling, we used anisotropic exchange interactions between nearest neighbors. The corresponding $\Gamma$-tensors taken from the DFT+$U$+SO calculations have diagonal form for all bonds and favor spin alignment along the $c$-axis.  

From Fig.\ref{im:Susceptibility} one can see that the magnetic susceptibility calculated using \textit{ab initio} microscopic parameters reproduces experimental data quite well. Here, we used the averaged value of the $g$-factor, $g$ = 2.138 estimated from the first-principles calculations  

\begin{figure}[!h]
\includegraphics[width=0.44\textwidth]{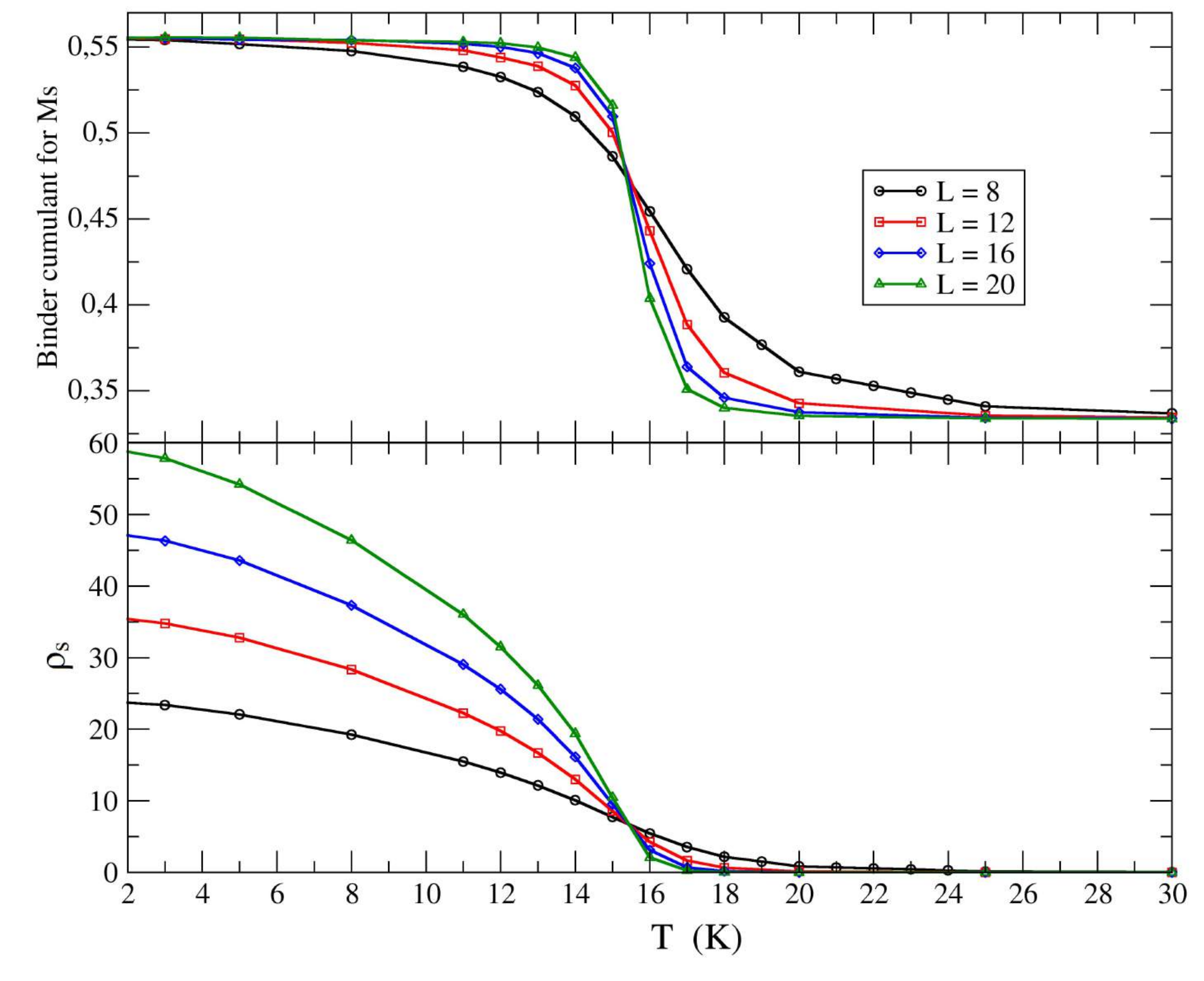}
\caption{Determination of the N\'eel temperature. (Top panel) Binder's cumulant for staggered magnetization; (Bottom panel) the calculated spin stiffness.  The crossing point for lattices with different linear sizes $L$ indicates the phase transition at $T_N=15.5$\,K.}
\label{im:Phase_transition}
\end{figure}

To calculate the N\'eel temperature, we used two complementary approaches proposed in Ref.~\onlinecite{Binder, stiffness}. They are based on the determination of the spin stiffness and Binder's cumulant for staggered magnetization at different temperatures and for different lattice size. These results are represented in Fig.~\ref{im:Phase_transition} (top and bottom panels, respectively). The crossing point for finite lattices of different size identifies the N\'eel temperature $T_N=15.5$\,K, which is in good agreement with the experimental value of 16.4\,K. 

\begin{figure}[!h]
\includegraphics[width=0.44\textwidth]{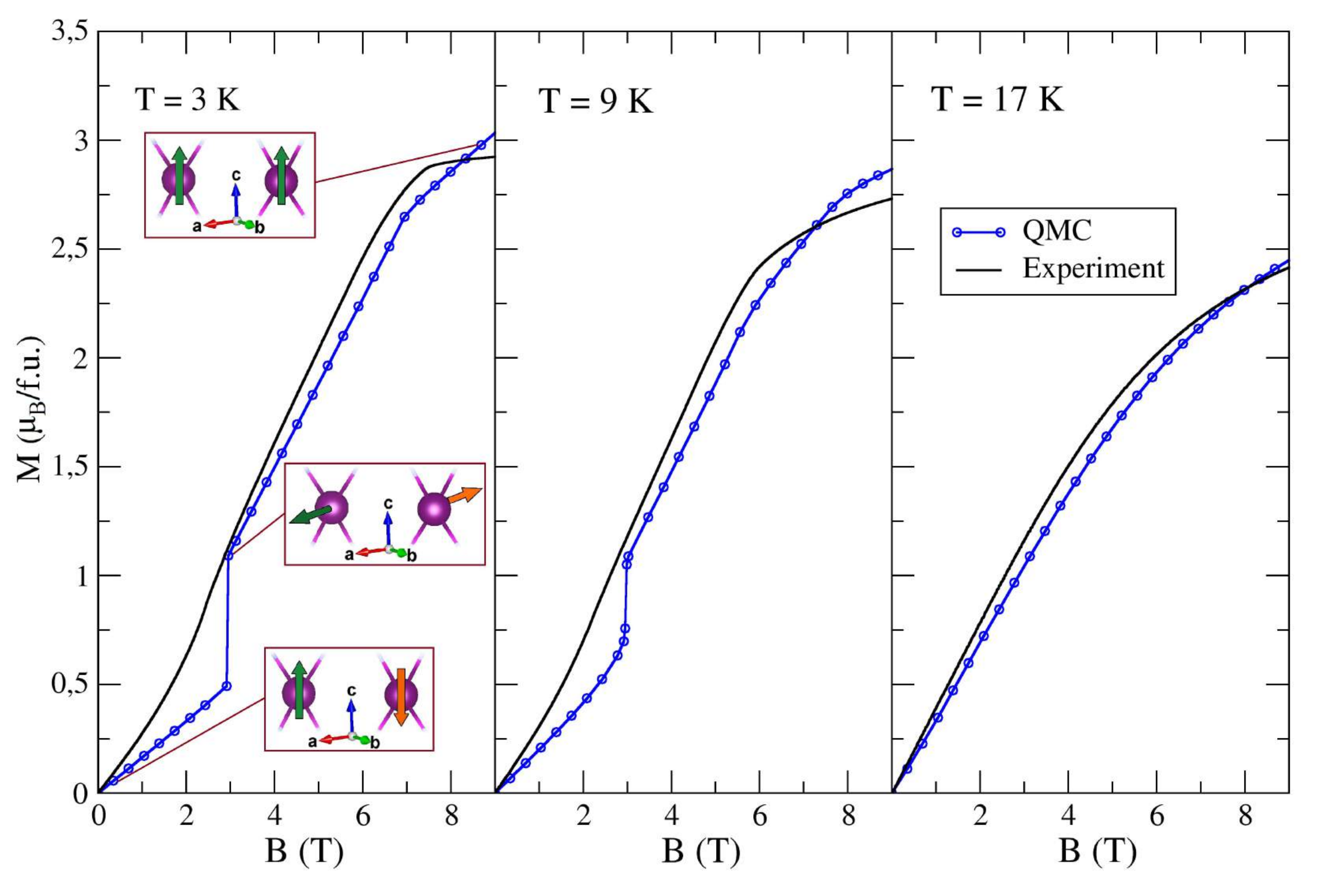}
\caption{Results of QMC simulation for magnetization at different temperatures. The insert shows the direction of local magnetic moments of copper atoms in nearest-neighbor chains. }
\label{im:Magnetization}
\end{figure}

Since the magnetization data were obtained for polycrystalline sample, we simulated magnetization of our anisotropic spin model for the $c$ and $a+b$ directions of the external magnetic field. The powder-averaged magnetization curve is presented in Fig.~\ref{im:Magnetization}. Within our approach we can reproduce the experimental profile of the magnetization at small magnetic fields that is characterized by the deviation from the linear behavior. This feature of the $M(B)$ curve originates from the anisotropic exchange interaction. In the insert of Fig.~\ref{im:Magnetization}, the directions of local magnetic moments in nearest-neighbor chains are shown. Upon increasing external magnetic field, the ground state goes through a spin-flop transition toward collinear ferromagnetic order between the chains.

Another important quantity characterizing a low-dimensional spin system is the value of the local magnetic moment. In quantum systems with $S=\frac{1}{2}$, it can be considerably reduced from its nominal value of 1\,$\mu_{B}$. For instance, in the case of the antiferromagnetic square lattice~\cite{Katanin1, Katanin2} the value of the magnetic moment equals to 0.6\,$\mu_B$. It indicates a significant suppression of the magnetic moment by quantum fluctuations in a low-dimensional antiferromagnetic spin system. 

For a realistic estimation of ordered magnetic moment in Ba$_3$Cu$_3$Sc$_4$O$_{12}$, we used the procedure proposed in Ref.~\onlinecite{Reger}. To this end, the non-local spin-spin correlation function $C_{L/2}$ between the most distant spins of a finite lattice is used. The corresponding value of the magnetic moment per site equals to $m = \lim \limits_{L\to\infty} \sqrt{3C_{L/2}}$. The ordered moment in the thermodynamic limit $L\to\infty$ can be estimated by the extrapolation procedure~\cite{Sandvik}:  
\begin{equation}
M^2(L) = 3C_{L/2} =  m^2 +\frac{n_1}{L}+\frac{n_2}{L^2}+\frac{n_3}{L^3},
\end{equation}
where $n_1$, $n_2$ and $n_3$ are constants. Because of two inequivalent positions of copper atoms in the  Ba$_{3}$Cu$_{3}$Sc$_{4}$O$_{12}$ system, we have calculated the spin-spin correlation functions for Cu$^{\rm I}$ and Cu$^{\rm II}$  sublattices, respectively. These simulations produce the local magnetic moments of 1~$\mu_{B}$ on both Cu sites (insert of Fig.~\ref{im:Susceptibility}). 

\subsection{\label{sec:level3}Neutron spectrum simulation}
\begin{figure}[!h]
\includegraphics[width=0.48\textwidth]{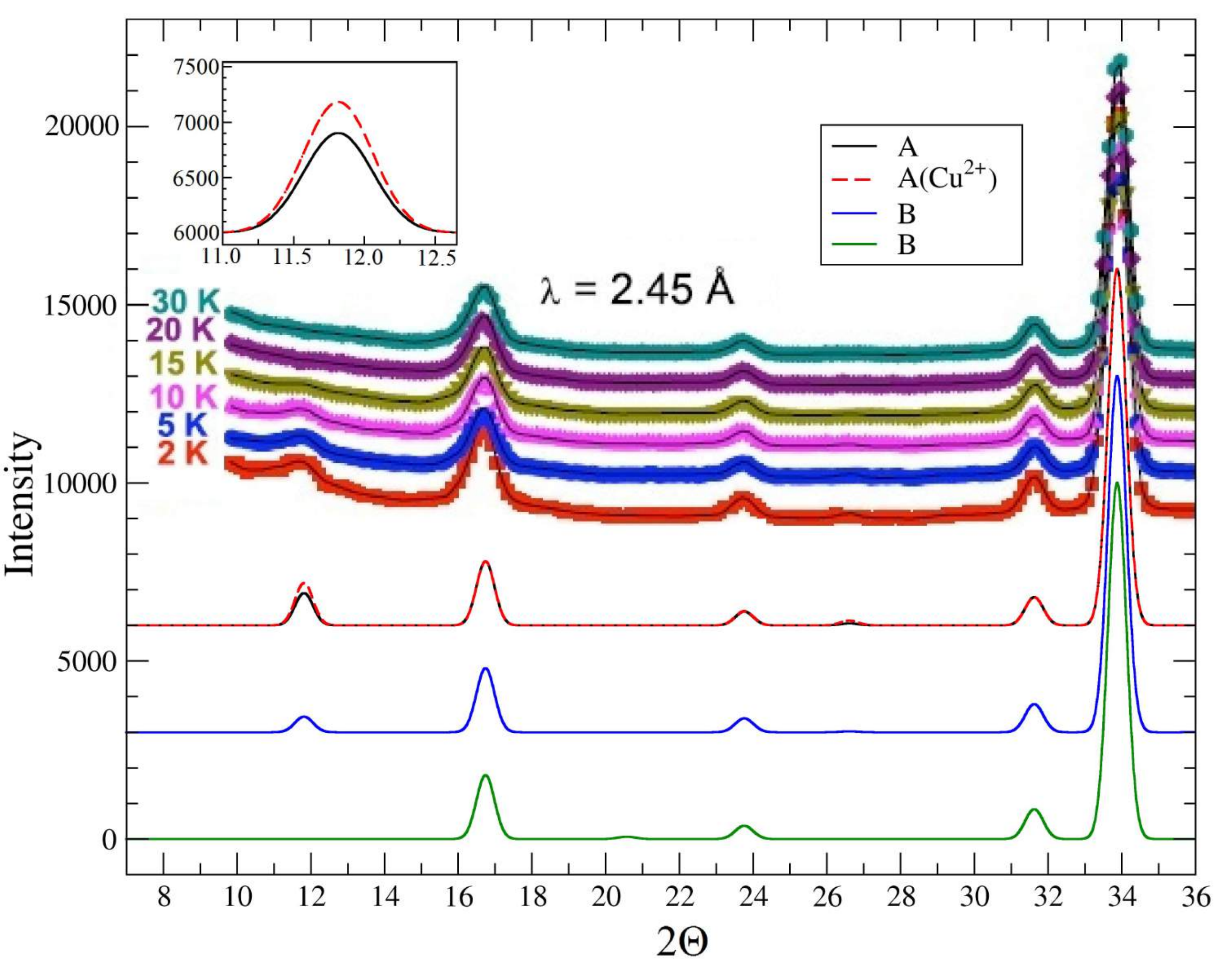}
\caption{Comparison of the experimental and theoretical neutron diffraction patterns taking into account the  hybridization effect on the magnetic form-factor. The experimental data (bold lines) are taken from Ref.~\onlinecite{Dasgupta}. Thin black, red, blue, and green lines denote the simulation results for different magnetic configurations presented in Fig.~\ref{im:Local_minimum}. The insert demonstrates the effect of the copper-oxygen hybridization on the neutron scattering.}
\label{im:Neutron}
\end{figure}

To confirm the proposed magnetic ground state of Ba$_3$Cu$_3$Sc$_4$O$_{12}$, we simulated neutron diffraction data for candidate spin configurations. Experimental results are taken from Ref.~\onlinecite{Dasgupta} and represented in Fig.~\ref{im:Neutron}. A weak low-angle magnetic reflection at 11.8$^\circ$ is clearly seen at 2\,K. This reflection disappears at about 20\,K, which is above the N\'eel temperature. It is also suppressed in magnetic fields of about 6\,T. While other magnetic reflections may be present as well, the experimental resolution was not high enough to detect them. 

For modeling neutron diffraction patterns we used the Jana2006 package~\cite{Jana}. Such simulations were performed for three magnetic configurations presented in Fig.\ref{im:Local_minimum}. The resulting spectra are represented in Fig.~\ref{im:Neutron}. The comparison of the neutron spectra obtained with ionic and covalent magnetic form factors (Fig.\ref{im:F-factor}) has shown that the account of the metal-ligand hybridization decreases the intensity of the magnetic reflection at 11.8$^\circ$. 

Experimental neutron diffraction data are compatible with both configurations $\mathbf{A}$ and $\mathbf{B}$. From the intensity ratio between the magnetic and nuclear peaks at 11.8$^\circ$ and 16.7$^{\circ}$, respectively, model $\mathbf{B}$ shows a slightly better agreement with the experiment. We note, however, that the intensity of the magnetic reflection is largely determined by the size of the ordered moment and by the magnetic form-factor. It is also important that configurations $\mathbf{A}$ and $\mathbf{B}$ are very close in energy, and their relative stability may depend on the DFT exchange-correlation functional and $\tilde U$ value. On the other hand, the configuration $\mathbf{C}$ proposed in Ref.~\onlinecite{Volkova} can be clearly discarded, because it reveals zero intensity of the magnetic reflection at 11.8$^\circ$.

\section{\label{sec:level1}Conclusions}
We have reported a comprehensive microscopic description of electronic and magnetic properties of Ba$_{3}$Cu$_{3}$Sc$_{4}$O$_{12}$, a spin-$\frac12$ low-dimensional magnet featuring distinct paper-chain topology of spin chains formed by CuO$_4$ plaquettes. We show that interactions within the paper chains are ferromagnetic and induce ferromagnetic order along the chains, which are ordered antiferromagnetically. By means of the superexchange theory and first-principles calculations, symmetric and antisymmetic anisotropic exchange interactions were determined. Quantum Monte Carlo similations for the resulting magnetic model reveal excellent agreement with the experimental N\'eel temperature, magnetic susceptibility, field-dependent magnetization, and elastic neutron scattering data. We demonstrate minor role of intrachain magnetic frustration and pinpoint fingerprints of the magnetic anisotropy in the non-linear behavior of magnetization as a function of field.

\section{\label{sec:level1}Acknowledgments}
The work of DIB, VVM and IVS is supported by the grant program of the Russian Science Foundation 14-12-00306. The numerical simulations were performed on the high-performance cluster of Ural Federal University. AAT was supported by the Federal Ministry for Education and Research through the Sofja Kovalevskaya Award of Alexander von Humboldt Foundation.

\end{document}